\documentstyle[spie]{article}  
\input{psfig}    

\newcommand{\um}{\,\hbox{$\mu$m}}
\def\deg{\nobreak{$^\circ$}}
\def\arcsec{\nobreak{$''$}}

\def\spose#1{\hbox to 0pt{#1\hss}}
\def\simlt{\mathrel{\spose{\lower 3pt\hbox{$\mathchar"218$}}
     \raise 2.0pt\hbox{$\mathchar"13C$}}}
\def\simgt{\mathrel{\spose{\lower 3pt\hbox{$\mathchar"218$}}
     \raise 2.0pt\hbox{$\mathchar"13E$}}}

\def\plotfiddle#1#2#3#4#5#6#7{\centering \leavevmode
    \vbox to#2{\rule{0pt}{#2}}
    \includegraphics{#1}}

\title{ALFA: First Operational Experience of the MPE/MPIA Laser Guide
Star System for Adaptive Optics}
 
\author{R. I. Davies\supit{a}, W. Hackenberg\supit{a},
T. Ott\supit{a}, A. Eckart\supit{a}, H.-C. Holstenberg\supit{a},\\
S. Rabien\supit{a}, A. Quirrenbach\supit{b} and
M. Kasper\supit{c}
\skiplinehalf  
\supit{a}Max-Planck-Institut f\"ur extraterrestrische Physik, 85740
Garching, Germany 
\skiplinehalf  
\supit{b}University of California, San Diego
\skiplinehalf
\supit{c}Max-Planck-Institut f\"ur Astronomie, K\"onigstuhl 17, 69117
Heidelberg, Germany 
\skiplinehalf  
See paper 3353-05 for other MPIA project members and report on the AO
system
} 
 
\authorinfo{Other information:\\
E-mail: davies@mpa-garching.mpg.de\\
ALFA homepage: http://www.mpia-hd.mpg.de/MPIA/Projects/ALFA/index.html}

\begin{document}  
  \maketitle  
 
\begin{abstract} 

The sodium laser guide star adaptive optics system ALFA has been
constructed at the Calar Alto 3.5-m telescope.
Following the first detection of the laser beacon on the wavefront
sensor in 1997 the system is now being optimized for best
performance.
In this contribution we discuss the current status of the launch beam
and the planned improvements and upgrades.
We report on the performance level achieved when it is used with the
adaptive optics system, and relate
various aspects of our experience during operation of the system.
We have begun to produce scientific results and mention two of these.

\end{abstract} 
 
 
 
\section{INTRODUCTION} 
\label{sect:intro}  

ALFA is a laser guide star adaptive optics system for the 3.5-m
telescope at the German-Spanish Centre for Astronomy on Calar Alto,
with a performance goal of achieving 50\% Strehl at
2.2\um\ under typical seeing conditions (0.9\arcsec) with good sky coverage.
It has been built as a joint venture
between the Max-Planck-Institut f\"ur Astronomie (Heidelberg)
responsible for the wavefront sensing and adaptive optics, and the
Max-Planck-Institut f\"ur extraterrestrische Physik (Garching)
who provide the laser guide star.

The AO system utilises a Shack-Hartmann sensor with several lenslet
arrays which can be interchanged, normally providing 6 or 18
subapertures (arranged in a hexagonal pattern)
which are sampled at rates of 60--900\,Hz.
The centroids of the spots in the sensor are used to determine the
coefficients for Zernike or Karhunen-Loeve modes, and these are then
applied to a 97-actuator mirror.
When correcting on a natural guide star, the wavefront sensor
determines all the modes including tip and tilt.
When using the laser guide star a separate tracker is used to
measure tip and tilt, since the up-link and down-link tip-tilt motions
associated with a laser beacon are correlated.
This part of the project is described in more detail in this
proceedings and elsewhere \cite{gli97,hip98}.

The optical bench for the laser is installed in the coud\'e lab, where
an Ar$^+$ laser with a 25\,W multiline output pumps a
dye ring laser with a single line output power of 4.25\,W. 
The lasers are continuous-wave to avoid difficulties of
synchronisation with the AO.
A detailed description of them can be found in other proceedings
\cite{qui97a,qui97b}.
The output beam is fed along the coud\'e path
and directed into a launch telescope.
At zenith the resulting sodium beacon has a magnitude of about m$_{\rm
V} = 9$--10.
Progress on optimizing the entire system has been hampered by poor
weather, but important milestones include closing the loop on the laser
beacon in Sept 1997, and then using it to enhance image resolution in
Dec 1997 when 6 subapertures were sampled at a rate of 60\,Hz,
allowing correction of 7 modes (plus tip and tilt) with a disturbance
rejection bandwidth of approximately 6\,Hz.
The latest result from March 1998, is that we have seen the laser guide
star on the wavefront sensor using 18 subapertures at a frame rate of
100\,Hz (Figure~\ref{fig:18ap}). 
We did not try to lock on it due to some focussing difficulties,
although the signal-to-noise ratio suggests that this should be
possible.

In this contribution we report on the current status of the laser
system, and discuss issues that have arisen from our operational
experience.
We describe proposed experiments and plans for the near future.
We have begun to produce scientific results, not only with the adaptive
optics, but using the laser beacon as a reference and we include a
section relating these.

\section{THE LAUNCH BEAM}   

\begin{figure}[ht]
\plotfiddle{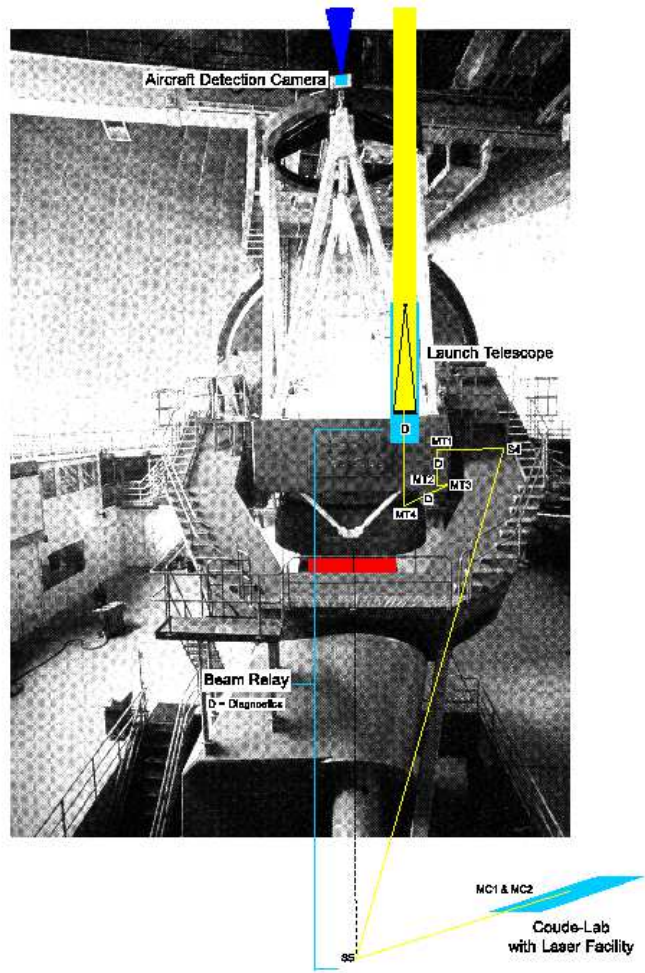}{15cm}{0}{63}{63}{-180}{-70}
\caption{The 3.5-m telescope at Calar Alto and the ALFA LGS-system.
The beam relay from the coud\'e lab is shown, with the position of
the launch telescope and the aircraft detection camera ALIENS.
Alignment of all the mirrors is achieved by use of diagnostic
cameras (denoted D) installed at critical points.}
\label{fig:35m}
\end{figure}

The design of the laser system incorporates an optical bench in the coud\'{e}
lab on which the Coherent INNOVA 200 Pump and Coherent 899 dye lasers
are installed, as there are facilities for providing the 50\,kW pump
power, cooling water, and dye circulation system.
Preliminary control features are included at this stage \cite{qui97a,qui97b}. 
Figure~\ref{fig:35m} shows how the output beam is directed by a
succession of mirrors along the coud\'e path to the telescope where
it is picked off near the primary mirror and directed into the
Cassegrain launch telescope.
There is another optical bench here where further diagnostic
instruments can be installed before the beam is projected upwards.

\subsection{Current Status}

There are a number of components on the optical bench which are designed
to optimize the quality of the launch beam, and hence the size and power
of the laser beacon.
These are described in the order in which the output beam from the dye
laser encounters them.

\subsubsection{Line-width}

A small fraction of the beam power is diverted to a reference cavity
which, for absolute wavelength tuning and long-term stabilisation, is
locked to the Lamb dip in the fluorescence signal from a sodium cell.
In principle this allows the laser line-width to be reduced to around
1\,MHz but in practice it is sufficient to keep it around 10\,MHz, the
natural line-width of the transition.
It also allows the laser to be de-tuned away from the Na line for
diagnostic purposes.

\subsubsection{Polarisation}

The $^2$S$_{1/2}$ ground state of the Na atom is split into a doublet
with a separation much greater than the laser line-width, and
the laser is tuned to excite atoms from the higher angular
momentum level ($F=2$).
Since they can decay back to either the $F=1$ or $F=2$ levels, after a
few cycles the whole population will reside in the $F=1$ level,
resulting in a significant loss of pump efficiency even though
the atoms are replenished by high altitude winds.
On the other hand, if the laser beam is circularly polarised so that it
imparts angular momentum to the atoms, they will tend to decay back
to the $F=2$ level, increasing the efficiency.

Since the output from the dye laser is plane-polarised, it can be
circularly polarised simply by including a $\lambda/4$-plate in the
path.
We expect a gain of 10--20\%, but with two stipulations.
The first is that between the coud\'e lab and the launch telescope
there are many reflections so that the final polarisation state of
the beam is uncertain, probably elliptical.
The second is that the polarisation only becomes an important issue if
the beam is close to saturating the Na layer, otherwise the loss of
atoms from the $F=2$ state is a minor effect.

Preliminary experiments suggest that currently the beam polarisation
does not significantly affect the beacon brightness, and that the low
power and large spot size are the major limitations.

\subsubsection{Beacon size}

The angular size of the laser beacon needs to be minimised so that
it is more intense and the centroid better defined.
Additional errors can arise with a LGS larger than the isoplanatic
patch at the wavelength of the science observations, particularly
if the size of the apertures on the wavefront sensor is well matched
to the coherence length $r_0$. 
However, this is not a serious problem for us, since our system is
designed to operate in the near-infrared where $r_0$ is relatively
large.

The projected beam is focussed on the Na layer from a Cassegrain launch
telescope.
There is a pre-expander on the optical bench which increases the beam
width to 1.5--2.5\,cm diameter, controlling the size of the beam
exiting the launch telescope within the range 29--49\,cm.
Two conflicting effects determine the diameter used:
the theoretical minimum spot size is reduced if a wider exit-beam
diameter is used due to diffraction effects; but once this exceeds
a few times the coherence length $r_0$, atmospheric turbulence will
act to increase the spot size.
In order to improve on this, the laser is optimized to operate in a
single (Gaussian) mode so that most of the power is concentrated in a
small region.

Typically we achieve a spot size of 3\arcsec, most of which is
probably a result of dome seeing and turbulence arising between the
coud\'e lab and telescope, for example due to temperature differences
between the coud\'{e} lab and dome.
To combat this we have already enclosed the entire path in a pipe, a
scheme that produced a noticeable improvment and also acts as a
safety precaution.

\subsubsection{Brightness}

The routinely obtainable laser output power is 4.25\,W.
However, this is significantly reduced to about 60\% by the time it
reaches the telescope due to the multiple reflections from mirrors
which very quickly become dirtied with dust blown over from the Sahara.

The addition of COT to the Rh6G glycol dye solution increases the laser
power by about 10\% by further suppressing the triplet states in the
dye molecules.
However, we have found that this lasts only for short time-scales
of tens of minutes.
A further effect is that the lifetime of the molecules is extended,
although the reason for this is unclear.

\begin{figure}
\plotfiddle{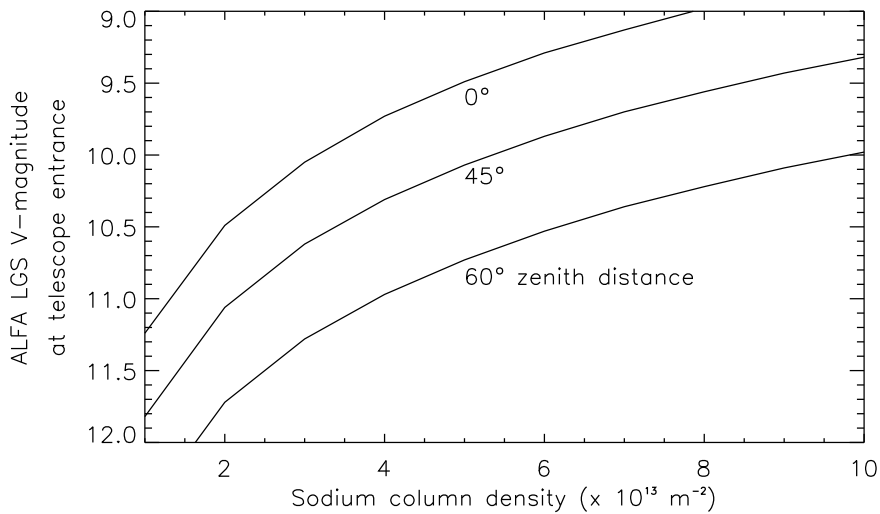}{6cm}{0}{85}{85}{-220}{-320}
\caption{Predicted LGS brightness (given as equivalent V-band
magnitude) as a function of zenith angle and Na column density,
assuming a total transmission for excitation (including atmosphere) of 50\%.}
\label{fig:lgsmag}
\end{figure}

The magnitude of the LGS has a strong dependence on zenith angle and
Na column density, and Figure~\ref{fig:lgsmag} shows the prediction of
how they affect the beacon magnitude.

\subsection{Operational Issues}

\subsubsection{Software}

The control software has been implemented on a set of industrial VMEbus 
computers and UNIX workstations and is significant for reliability, usability
and maintainability of the whole system. 
It uses EPICS for distributed realtime interprocess communication.

\subsubsection{Focus and alignment}

Two important issues which must be considered throughout any
observation are the focus and alignment of the beam as these both
depend on zenith angle due to the change in distance to the sodium layer,
the variable alignment being a result of the offset between the axis of
the launch beam and that of the telescope.
A further consequence of the varying zenith angle is that the adaptive
optics must be re-calibrated at intervals in order to suppress the
focus term that would soon dominate.

Currently focus and alignment are both achieved manually using the
telescope's TV finder.
The alignment is then fine-tuned on the wavefront sensor so that the
spots seen through the lenslet array on the Shack-Hartmann sensor are
on average centred in the regions in which the AO algorithms
operate.
Once centred, the control software is able to maintain the correct
orientation of the mirrors independent
of the telescope's motion (a far from trivial task) by use of 3
additional pilot beams and 2 control loops for beam centring and
pointing in the upper and lower sections of the beam relay.

As mentioned previously, a small intense spot is required for the
centroiding algorithm to work optimally.
Our typical spot size of 3\arcsec\ is near the maximum size on which it
is possible to usefully correct, and attempts are being made to improve
on this situation.

\subsubsection{ALIENS}

A high power laser beam can pose a significant hazard to planes and
satellites which cross its path and it is essential to include certain
safety measures while the laser is operating.
ALIENS (Aircraft Light Imaging Emergency Notifcation System) is a CCD
camera with a 20\deg\ field of view, operating in the visual waveband.
It is mounted behind the secondary mirror of the main telescope to
insure that it always looks in the direction that the laser is
pointing.
It compares consecutive frames taken with an integration time
of 0.5\,sec to identify moving objects as changes
that exceed a specified threshold, and 
on a positive detection, a shutter closes off the laser beam.
Stars are automatically masked out as
scintillation can trigger a false alarm.

Following significant modifications to the software, ALIENS is now
in operation.
Obtaining statistics on its effectiveness and reliability is slow simply
because very few planes fly over Calar Alto; but it has responded to
satellites and the false detection rate is typically only once every
6\,hours.
Difficulties still arise from high speed clouds, and from scattered
light when the dome is moving.

An issue which has not been addressed concerns the possibility that the
Rayleigh back-scatter may lie in the field being observed by other
telescopes at the site, perhaps seriously compromising their
observations.

\section{PLANS FOR THE NEAR FUTURE}

Once we are able to minimise the size of the beacon, polarisation will
become an important issue.
We are therefore installing an instrument to measure the polarisation at
the launch telescope, together with a control loop fed back to the
$\lambda/4$-plate on the optical bench.
We will also determine whether the polarisation changes with telescope
position as a result of the different reflection angles.

Optimal focussing of the laser is extremely important, and using the TV
guider to achieve this is clearly unsatisfactory.
We propose to use the wavefront sensor, which is the most sensitive
camera available to us, for this purpose by creating an optical path in the
adaptive optics bench which bypasses the lenslet array so that a single
spot is seen on the Shack-Hartmann sensor.

We have discussed the problems associated with using mirrors to direct
the laser to the telescope.
As an alternative we are experimenting with using a fibre for the same
purpose, resulting both in a smaller power loss than we are currently
experiencing from dirty mirrors, and also avoiding the complexity
of their alignment.
However, this presents a new set of difficulties.
Experiments so far have been with a fused silica fibre which does not
preserve the polarisation, but it is possible to buy fibres which do, or
to reset the polarisation at the launch telescope.
In order to retain the mono-mode quality of the beam it is necessary to
use a fibre with a core diameter of only 4\um, but this has knock-on
effects such as increasing the Brillouin backscatter to as much as
50\%, and we have measured a backscatter of 1.7\,W with an input power
of 3.5\,W.
It arises from the build-up of standing acoustic waves
when high powers are forced into a narrow fibre, but
can be suppressed by phase modulating the input beam at a
frequency of about 100\,MHz.

In a similar vein we are proposing to install a Raman-fibre laser
during 1998/1999.
In order to be competitive with our current power, the intended output
will be at least 10\,W.
The difference is due to the broader line-width (2\,GHz) of the fibre
laser, so that most of the power is at wavelengths with a much smaller
sodium absorption cross-section (the doppler broadened Na profile is
about 2\,GHz).

\section{EXPERIMENTS AT CALAR ALTO}

\subsection{Tip-Tilt from a Laser Beacon}

The tip and tilt modes of a wavefront distortion are responsible for
the bulk of the image degradation.
Although this means that image resolution can be enhanced significantly
simply by use of a fast guider, it also has implications for laser guide
stars because the up-link and down-link tip-tilt components are
effectively the same.
Thus a laser beacon cannot easily be used to measure tip-tilt modes,
with the result that sky coverage is limited by the availability of
stars for fast guiding.
There are two alternative proposals to circumvent this difficulty.
One involves the polychromatic LGS concept \cite{foy95} in which the
4P$_{3/2}$ level
of Na is excited, and the ensuing radiative decay results in a line
spectrum from 0.33\, nm to 2.207\,\um. 
The wavelength dependence of the refractive index of air allows the
tip-tilt component to be determined.
The other method is to use auxiliary telescopes \cite{rag95} from which
the up-link and down-link motions of the laser and beacon are
uncorrelated.
The uncorrelated tip-tilt motions of the beacon and a natural guide
star are due to the up-link tip-tilt of the laser alone.
The advantage is that there is a much greater chance of finding a
natural star in the same isoplanatic patch as the laser, since it now
appears highly elongated.

Calar Alto has two telescopes which can be used to test the
latter method.
In collaboration with the TMR LGS network\footnote{the Network for Laser
Guide Stars on 8-m class Telescopes operates under the auspices of the
Training and Mobility of Researchers programme of the European Union}
during 1998 we will attempt the first detection of absolute tip-tilt
from a LGS system.

\subsection{Site Characteristics}
\label{sec:sitechar}

It is very important to have some knowledge of the atmospheric
characterisitcs of the site if successful and optimal adaptive optics
correction is to be achieved.
To this end, we propose to carry out a number of experiments in
collaboration with the TMR LGS network during the summer of 1998.

\subsubsection{Sodium layer profile}

\begin{figure}
\plotfiddle{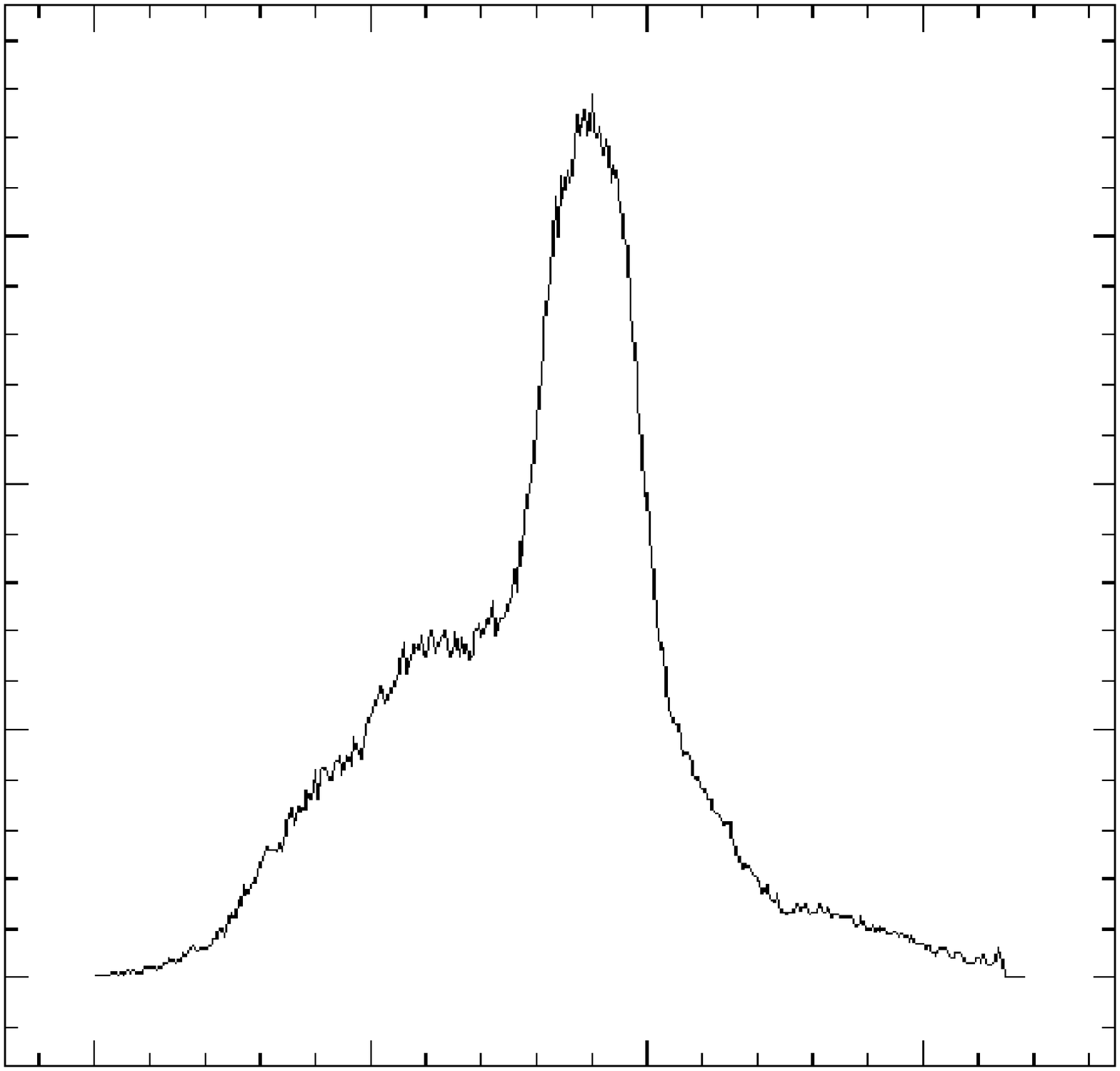}{6cm}{0}{30}{30}{-200}{-55}
\plotfiddle{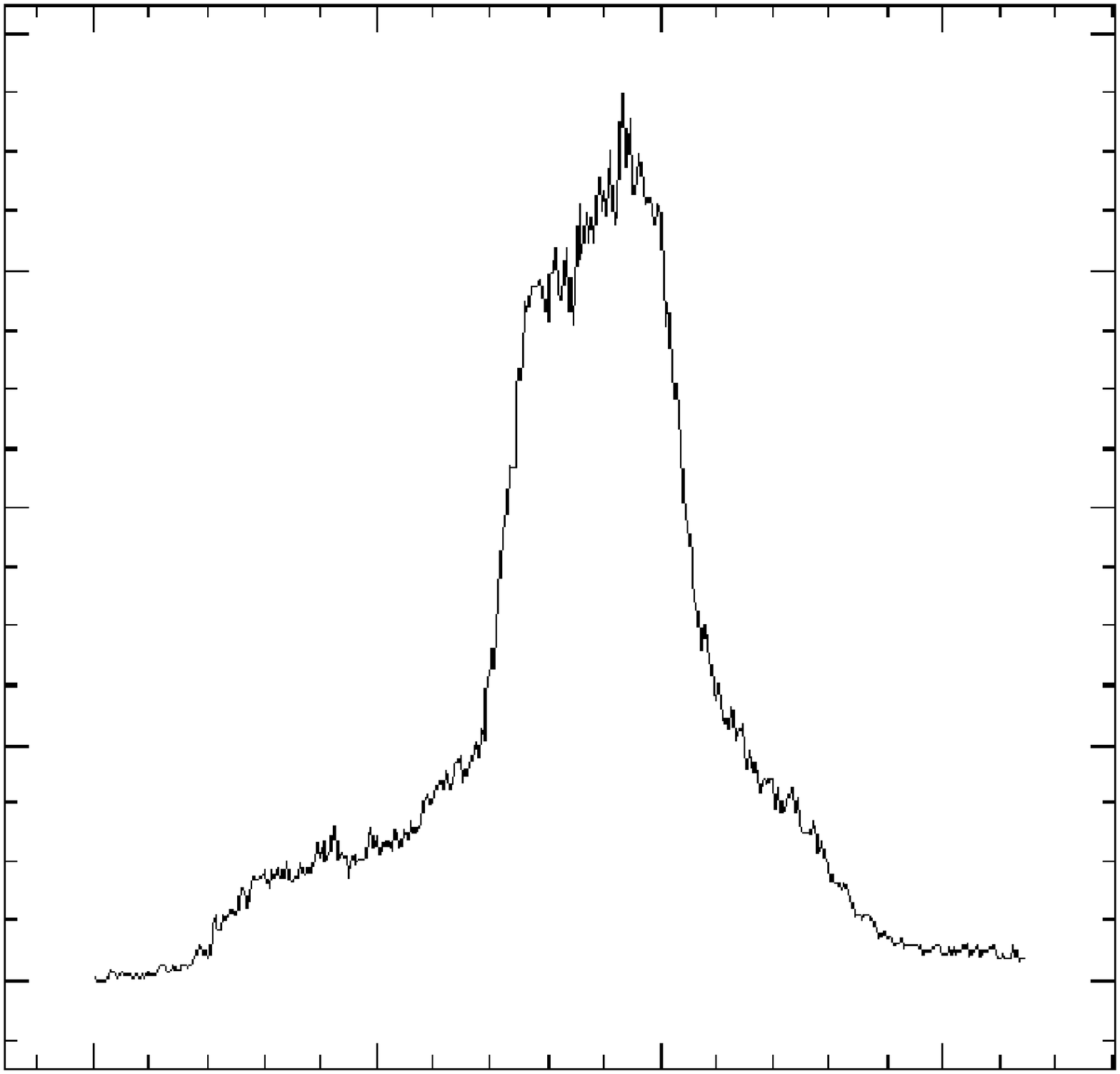}{0cm}{0}{29}{29}{+25}{-27}
\caption{Profiles of the ALFA LGS plumes as seen from the 1.2-m
telescope about 1\,hour apart, effectively slices through the Na layer.
The horizontal axis is distance along the plume, and the vertical axis
is intensity.
Changes in the Na layer profile are clearly seen.
The figure has kindly been provided by C. O'Sullivan and M. Redfern
(University College, Galway).}
\label{fig:plumes}
\end{figure}

At the Calar Alto site the 1.2-m and 2.2-m telesopes can be used to
observe the sodium plume of the ALFA laser, which appears from them as a
strip about 100\arcsec\ long.
Previous observations during the autumn of 1997 have shown that over
the time-scale of 1\,hour the profile of the Na layer can change
significantly, as shown in Figure~\ref{fig:plumes}
It is crucial to know the time-scale on which these structures appear
and disappear, and the scale of the differences.
One consequence is that a varying Na column density will affect the
brightness of the laser beacon;
another is that it may be necessary to refocus the launch beam, and also
the wavefront sensor, on short timescales.

\subsubsection{Atmospheric turbulence profile}

\begin{figure}
\plotfiddle{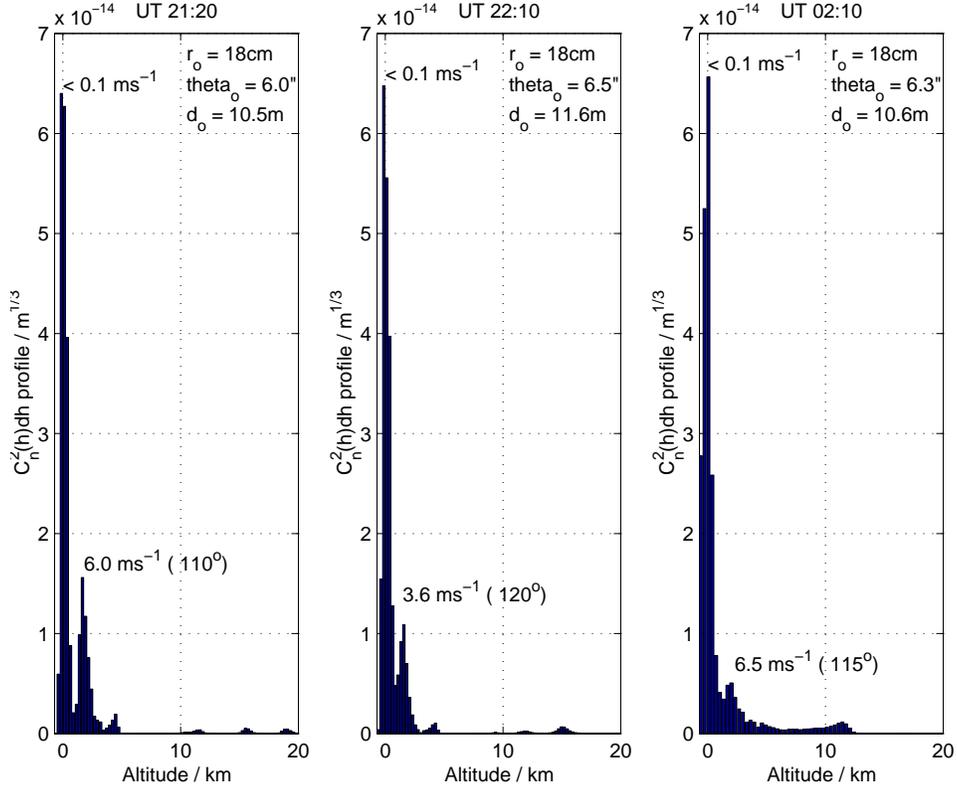}{10.2cm}{0}{75}{75}{-230}{-155}
\caption{SCIDAR results from Calar Alto at 3 times during a single
night.
The profile of the refractive index structure function C$_n^2$(h) is
shown, together with wind speed and direction for the dominant layers.
The coherence length $r_0$, isopanatic angle $\theta_0$, and LGS focal
anisoplanatic measure $d_0$, are calculated for $\lambda = 500$\,nm.
The figure has kindly been provided by V. Kl\"uckers and C. Dainty
(Imperial College, London)}

\label{fig:scidar}
\end{figure}

The effectiveness of any given AO configuration (sampling frequency,
number of modes corrected, etc.) depends on the prevailing atmospheric
conditions at the time.
A more practical approach is to measure the average turbulence and
velocity profiles at a particular site and time of year to gain an
understanding of the isoplanicity and time-scale for temporal variation.
The SCIDAR (SCIntillation Detection and Ranging) technique involves
measuring the spatial correlation of the scintillation patterns due to
a binary star in the telescope pupil plane in order to recover the
refractive index structure constant C$_n^2$(h) as a function of height.
The method also allows the wind speed and direction of the dominant
layers to be established.
Some data from one night in 1997 shown in
Figure~\ref{fig:scidar} suggest $r_0 = 18$\,cm in the optical, an
unexpectedly favourable result given the median seeing; more measurements
are needed to determine the typical characteristics.

\section{RESULTS}

The performance of the adaptive optics is integral to that of the laser
beacon, and during February 1998 the it was tested extensively using a
variety of subapertures, sampling frequencies, modes, and improved
centroiding algorithms.
Two particular examples are the achievement of an H-band FWHM of
0.24\arcsec\ on the m$_{\rm V} = 9.7$ star T\,Tau, operating at 200\,Hz
correcting 24 modes from 18 subapertures;
and the ability to close the loop on the m$_{\rm V} = 12.3$ star
GM\,Aurigae while correcting 7 modes from 6 subapertures at 60\,Hz.

\begin{figure}
\plotfiddle{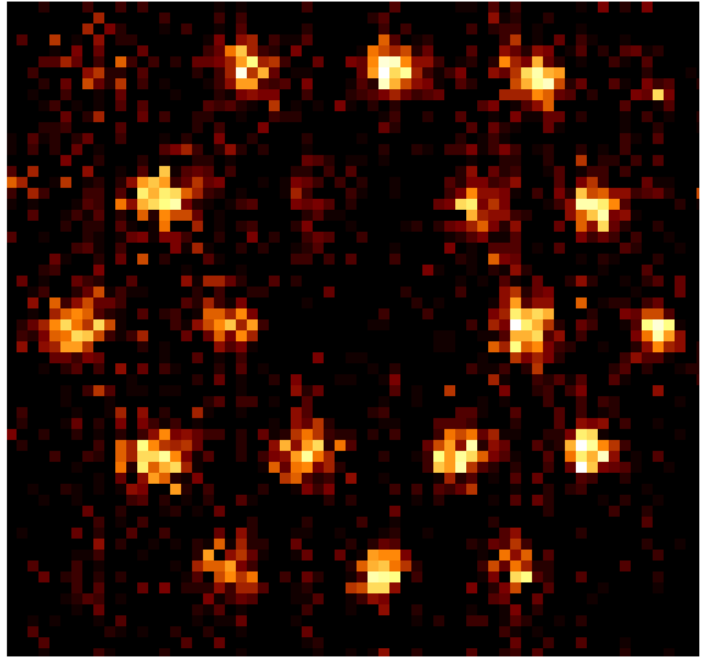}{9cm}{0}{60}{60}{-175}{-100}
\caption{ALFA Laser guide star as seen on the wavefront sensor using
the $5 \times 5$ lenslet array and sampling at 100\,Hz.
Some problems remain with focussing and we did not attempt to lock on
to it.}
\label{fig:18ap}
\end{figure}

These results indicate that the AO is at a stage where routine
correction using a laser guide star should be possible.
The detection in March 1998 of the laser on the wavefront sensor
through 18 subapertures at a frame rate of 100\,Hz, as shown in
Figure~\ref{fig:18ap}, is another indication of the progress being made.
In this case we did not attempt to lock on to it due to focussing
problems, but the count rate is close to that at which this should be
feasible.

\subsection{The Binary System BD\,+31\deg643}

A promising result from December 1997 is the first image improvement
attained by correcting on the laser guide star.
A K-band image of the 0.6\arcsec\ binary BD\,+31\deg643 (both B5
stars) in 
1.2\arcsec\ seeing appears as a single elongated object, but correcting
7 modes on the LGS at 60\,Hz (plus tip-tilt closed on the object at
100\,Hz) revealed the two stars clearly separated and with FHWM about
0.56\arcsec.
BD\,+31\deg643 has a dusty disk \cite{kal97}, and infact represents
only the second detection of a circumstellar dust disk around main
sequence stars, the other being around a single A0 star ($\beta$\,Pic).
At a distance of 330\,pc high spatial resolution is essential for
interpreting observations and furthering our understanding of dust disks
around main-sequence stars, a poorly studied phase in the
formation of planetesimals.
In this case it indicates that binary stars may possess a stable
environment for planetesimal formation.

\begin{figure}
\plotfiddle{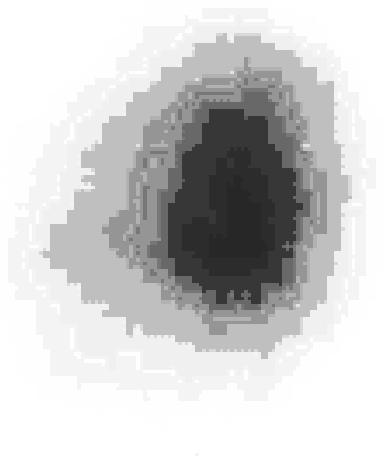}{4cm}{0}{90}{90}{-355}{-350}
\plotfiddle{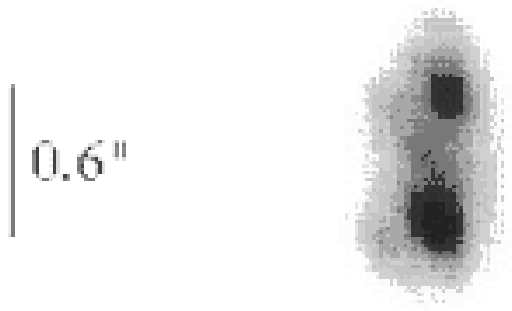}{0cm}{0}{90}{90}{-190}{-340}
\caption{K-band images of BD\,+31\deg643. Left is open-loop in seeing
of 1.2\,arcsec, and right is with the loop closed on the laser guide
star to achieve a resolution of 0.56\,arcsec (see text for details).}
\label{fig:bd31643}
\end{figure}

\subsection{Galaxies in the Abell 1367 and 262 Clusters}

Also observed in December~1997 and February~1998 were 25 galaxies in
the Abell\,1367 and Abell\,262 clusters, including UGC\,1347 and
UGC\,1344 (in preparation).
We used a close ($\sim$30\arcsec) natural guide star,
 again correcting 7 modes at a sampling rate of 60\,Hz,
achieving a disturbance rejection bandwidth of about 5\,Hz.
Comparison of the closed-loop data with open-loop data taken immediately
afterwards showed that the resolution had been improved from 1.2\arcsec\
to 0.9\arcsec.
For the latter we corrected on a nearby star through the same number of
subapertures at a higher frequency (giving a slightly higher rejection
bandwidth) and increasing the resolution in this case from 0.9\arcsec\ to
0.6\arcsec.
A second star in the same field as UGC\,1347 and at approximately the same
separation from the reference star could be used as a check, showing that
all the sources were well within the isoplanatic patch and that the
reference stars represented an accurate portrayal of the point
spread function to clean the images.

The detailed study of these two galaxies showed that there is evidence
for nuclear star formation, and that part of the nuclear K-band
luminosity can be explained by the presence of a young stellar population.
The larger sample of galaxies indicates a correlation between
the size of the galaxy bulge in the K-band, and the distance from the
centre of the cluster. 
This is explained in terms of enhanced star formation in gas-rich
spirals triggered by galaxy interactions as they cross the central
regions of the cluster.
The additional compact component of flux from this activity
then influences the apparent size of the bulge.
The typical bulge size is $\simlt 1$\arcsec\ ($\simlt 300$\,pc), and
high spatial resolution is essential to measure this accurately
as provided by adaptive optics with use of a LGS.


\acknowledgments     
  
The MPIA/MPE team thanks the Calar Alto staff for their help and
hospitality, and N. Wilnhammer for technical support.
RID acknowledges the support of the TMR (Training and Mobility of
Researchers) programme as part of the European Network for Laser Guide
Stars on 8-m Class Telescopes.
We thank Mike Redfern and Creidhe O'Sullivan (University College, Galway),
and Vince Kl\"uckers and Chris Dainty (Imperial College, London) who
provided the figures for Section~\ref{sec:sitechar}.

 
  \bibliography{kona_paper}   
  \bibliographystyle{../kona/spiebib}   
  
  \end{document}